\documentclass[useAMS,usenatbib]{mn2e}

\usepackage{graphicx,amssymb,bm}
\usepackage{subfig,xfrac}
\usepackage{float, Abbrev}
\usepackage[fleqn]{amsmath}  
\usepackage{color}
\usepackage[draft]{hyperref}

\usepackage{multirow}

\newcommand{\be}{\begin{equation}}
\newcommand{\ee}{\end{equation}}
\newcommand{\ba}{\begin{eqnarray}}
\newcommand{\ea}{\end{eqnarray}}

\def\simlt{\lower.5ex\hbox{$\; \buildrel < \over \sim \;$}}

\newcommand{\fig}{\begin{figure} \begin{center}}
\newcommand{\efig}{\end{center}\end{figure} }
\newcommand{\figs}{\begin{figure*}\begin{minipage}{180mm} \begin{center}}
\newcommand{\efigs}{\end{center}\end{minipage}\end{figure*} }
\def\simgt{\lower.5ex\hbox{$\; \buildrel > \over \sim \;$}}

\usepackage{graphicx} 

\title[Constrain SIDM with BCGs]{Observable tests of self-interacting dark matter in galaxy clusters: BCG wobbles in a constant density core}
\author[D. Harvey et al]
{David Harvey$^{1,2}$\thanks{e-mail: {\tt david.harvey@epfl.ch}}, Andrew Robertson$^3$, Richard Massey $^3$ and Ian G. McCarthy$^4$ \\
$^{1}$Lorentz Institute, Leiden University, Niels Bohrweg 2, Leiden, NL-2333 CA, The Netherlands \\
$^{2}$Laboratoire d'Astrophysique, EPFL, Observatoire de Sauverny, 1290 Versoix, Switzerland \\
$^{3}$Institute for Computational Cosmology, Durham University, South Road, Durham, DH1 3LE, UK \\
$^{4}$Astrophysics Research Institute, Liverpool John Moores University, 146 Brownlow Hill, Liverpool L3 5RF}

\begin{document}

\date{Accepted ---. Received ---; in original form \today.}

\pagerange{\pageref{firstpage}--\pageref{lastpage}} \pubyear{2017}

\maketitle

\label{firstpage}

\begin{abstract}

Models of Cold Dark Matter predict that the distribution of dark matter in galaxy clusters should be cuspy, centrally concentrated.
Constant density cores would be strong evidence for beyond-CDM physics, such as Self-Interacting Dark Matter (SIDM). % with non-zero cross-section to non-gravitational forces.
An observable consequence would be oscillations of the Brightest Cluster Galaxy (BCG) in otherwise relaxed galaxy clusters.
Offset BCGs have indeed been observed -- but only interpreted via a simplified, analytic model of oscillations.
We compare these observations to the BAHAMAS-SIDM suite of cosmological simulations, which include SIDM and a fully hydrodynamical treatment of star formation and feedback.
We predict that the median offset of BCGs increases with the SIDM cross-section, cluster mass and the amount of stellar mass within $10$kpc, while CDM exhibits no trend in mass.
Interpolating between the simulated cross-sections, we find that the observations (of 10 clusters) is  consistent with CDM at the $\sim1.5\sigma$ level, and prefer cross-section $\sigma/m<0.12 (0.39) $cm$^2$/g at 68\% (95\%) confidence level.
%This is on the verge of discriminating between velocity-independent dark matter models that would explain discrepancies in the behaviour of dwarf galaxies, and will be improved by larger surveys by Euclid or SuperBIT.
This is on the verge of ruling out velocity-independent dark matter self-interactions as the solution to discrepancies between the predicted and observed behaviour of dwarf galaxies, and will be improved by larger surveys by Euclid or SuperBIT.

%Idealised simulations suggest that the oscillations of a Brightest Cluster Galaxy (BCG) within a massive galaxy cluster is a unique signature of a cored density profile and potentially beyond Cold Dark Matter physics.
%A subsequent observational study claimed to have observed this, finding a $3\sigma$ discrepancy with CDM. 
%We use full hydrodynamical cosmological simulations of self-interacting dark matter to test these claims.
%Using the BAHAMAS suite of simulations that include four velocity independent and elastic cross-sections ($\sigma/m=0,0.1,0.3,1.0$cm$^2$/g) plus a full baryonic feedback prescription, we measure the theoretical distributions of offsets between dark matter and the central BCG. 
%We model these offsets as a log-normal distribution, and find that for increasing non-zero cross-sections the linear trend between the log-normal median, $\mu$ and halo mass increases, while CDM exhibits no trend in mass.
%We interpolate between the cross-sections to create a model that predicts $\mu$ as a function of cross-section and halo mass. 
%Using this model we find the data prefers a cross-section of  $\sigma/m<0.22$cm$^2$/g (95\% CL) with 38\% probability that they are consistent with CDM. This finding challenges a velocity independent self-interacting dark matter that can have an observational impact on dwarf galaxy scales.
%We find that this method has potential in our efforts to constrain this fundamental parameter for upcoming missions such as Euclid and SuperBIT, for both weak and strong gravitational lensing.

\end{abstract}


\section{Introduction}
The search for dark matter remains fruitless. As the dominant mass component in our Universe, revealing its nature has become one of the greatest questions of modern science.  However, despite wide efforts to detect it, for example at the Large Hadron Collider in Cern \citep{LHC_review}, or directly at the LUX experiment\citep{LUX_2016} the community remains in the dark.

In an effort to diversify and broaden our search, physicists have begun to consider new avenues, focusing on specific properties of dark matter.
In this paper we address one such property, the self-interaction cross-section. Dark matter is commonly assumed to be collisionless. However, dark matter that exhibits a relatively large self-interaction cross-section ($\sigma_{\rm DM} / m \gtrsim 0.5 \, $cm$^2$/g or 0.2 barn/GeV) could potentially alleviate problems that exist in the small-scale structure of the standard Cold Dark Matter model (CDM). By reducing the central densities of dark matter haloes and thus creating a core, it can ease the so-called core-cusp problem (where observations of dwarf galaxies suggest the existence of cored density profiles where simulations of CDM predict cuspy ones)\citep{corecusp,HaloSIDM,HaloSIDMA,HaloSIDMB,SIDMSimA}. It remains unclear whether these inconsistencies are due to unknown baryonic processes or a breakdown in the CDM model. However it is clear that by constraining SIDM we can rule it out as a cause of the small-scale problems, or probe self-interactions in the dark sector, something that is impossible with traditional dark matter experiments.

Efforts to constrain the momentum transfer cross-section per unit mass, $\sigma_{\rm DM} / m$ have been concentrated mainly on clusters of galaxies. Although some studies have looked at using dwarf galaxies \citep{SubhalosSIDMA,SIDM_FIRE,CoreFormationDwarf}, it remains to be seen if these observables are completely discriminative \citep{Harvey_dwarf,scultporModel}. Galaxy clusters, on the other hand, are favourable laboratories in which to probe dark matter self-interactions. The existence of large quantities of dark matter results in strongly deformed spacetime meaning that both strong and weak gravitational lensing can be used to infer its distribution.

Methods that use clusters of galaxies to constrain $\sigma_{\rm DM}/m$ can be classified into two distinct cases, those using merging clusters and those using relaxed ones. Although initially used due to their apparent simplicity, studies using relaxed clusters suffered from the lack of  high-resolution simulations, and hence found it difficult to place reliable constraints \citep[e.g.][]{SIDMTest}. As a result, in the past decade attention shifted to merging clusters. By comparing the distribution of dark matter to the collisionless galaxies many studies attempted to constrain the self-interaction cross-section to $\sigma_{\rm DM}/m \lesssim 1$\,cm$^2$/g \citep{impactpars,bulletcluster,Harvey15}.  However, subsequent studies have shown that uncertainties associated with the modelling and measurement interpretation can bias constraints \citep{SIDM_bullet,mismeasure}. It seems that the complex nature of these clusters means that gaining insightful conclusions will require high resolution simulations and careful modelling. 

The key observable that this paper will concentrate on was first proposed by \citet[][hereafter K17]{darkgiants}. They found that during the collision of two equal-mass clusters with cored density profiles, the Brightest Cluster Galaxy (BCG) would become offset from the centre of the halo. A constant central density leads to a gravitational potential that is quadratic in radius. An offset BCG therefore experience a harmonic oscillation long after the halo has re-relaxed and virialised. It was hypothesised that this observation would not be observed in CDM since the cuspy central region would keep the BCG tightly bound to the centre. 

Following this study, an observational paper looking at ten relaxed galaxy clusters attempted to observe this wobble \citep[][hereafter H17]{Harvey_BCG}. They used the parametric gravitational lensing algorithm {\sc Lenstool} to measure the positions of cluster-scale dark matter haloes from the locations of multiply-imaged background galaxies, and then measured the separations between the dark matter haloes and their corresponding BCGs. H17 found a wobble of $A_{\rm w}=11.8^{+7.2}_{-3.0}$kpc, where $A_{\rm w}$ is the amplitude of a harmonic oscillator that parameterises the distribution of dark matter--BCG offsets. Indeed when compared to {\it n}-body simulations, which included realistic baryonic feedback, there appeared to be a $3\sigma$ discrepancy with simulations predicting little or no wobble.

Due to a lack of SIDM simulations, H17 were unable to test for systematics associated with the harmonic oscillator model they used to model BCG wobbling. Moreover, the predictions of offset BCGs in K17 were from idealised, dark matter only simulations of equal mass mergers, not cosmological simulations of relaxed clusters. In this paper we build on these two studies by using cosmological simulations including baryonic physics of both CDM and SIDM, allowing us to characterise the BCG wobbling signal expected with CDM or with different SIDM models. 

This paper is structured as follows. In section \ref{sec:simulations} we outline the data used, including a recap of the H17 sample, the suite of simulations used, and how we select samples of simulated clusters. The next section outlines how we analyse these samples and we construct our model of the signal. In section \ref{sec:results} we fit our model, presenting our results, and in section \ref{sec:conc} we discuss our results and give our conclusions.

%%%%%%%%%%%%%%%%%%%%%%%%%%%%%%%%%%%%%%%%%%%%%%%%%%%%%%%%%%%%
%%%%%%%%%%%%%%%%%%%%%%%%%%%%%%%%%%%%%%%%%%%%%%%%%%%%%%%%%%%%
\begin{table}
\centering
\begin{tabular}{|c|c|c|c|c|}
Sample &$\sigma_{DM}/m$ (cm$^2$/g) &$N_{\rm cl}$ & $N_{\rm eff}$ &$\langle \log(M_{\rm tot}/M_{\odot})\rangle$ \\
\hline
CDM & 0.0 & 1365 &  460 & 14.45 \\ 
SIDM0.1 & 0.1 & 1344 &  731 & 14.32 \\ 
SIDM0.3 & 0.3 & 1374 &  672 & 14.42 \\ 
SIDM1 & 1.0 & 1330 &  645 & 14.40 \\ 
%CDMhires & 0.0 &   80 &   13 & 14.27 \\ 
%SIDM1hires & 1.0 &   95 &   18 & 14.25 \\ 
obs & N/A&   10 &   10 & 15.08 \\ 
\hline
\end{tabular}
\caption{The sample selection of galaxy clusters from the simulations with their corresponding dark matter cross-section. The third column gives the total number of clusters extracted from the simulation and the fourth column gives the number of relaxed clusters after cuts. The final column gives the mean halo mass of the cut sample.
\label{tab:data}}
\end{table}
%%%%%%%%%%%%%%%%%%%%%%%%%%%%%%%%%%%%%%%%%%%%%%%%%%%%%%%%%%%%
%%%%%%%%%%%%%%%%%%%%%%%%%%%%%%%%%%%%%%%%%%%%%%%%%%%%%%%%%%%%

\section{Data}{\label{sec:simulations}

\subsection{Observations}\label{sec:convolve}
In this paper we will use the observations from H17 in an attempt to measure the self-interaction cross-section of dark matter. 
 H17 looked at ten massive galaxy clusters ($\bar{z}=0.33$), with at least ten multiple images sourced from the Local Cluster Substructure Survey \citep{Locuss_Richard} and the Cluster Lensing And Supernova survey with Hubble \citep{CLASH_zitrin}. Using fitted parametric models of the main cluster halo and BCG, they used strong gravitational lensing to estimate the offset between the two in the plane of the sky. 
 
 In order to quantify the measurement uncertainty in the positioning due to a finite number of strong lensing constraints, they took the observed multiple images, derived source positions, then using a known model, projected these sources back in to the image plane. Using this new set of multiple images they measured the variance in the estimate of the best fit model, finding an RMS error of $\sigma_{\rm obs}=3.1$kpc. In this paper we will adopt the offsets observed along with its associated error estimate.
 
  %%%%%%%%%%%%%%%%%%%%%%%%%%%%%%%%%%%%%%%%%%%%%%%%%%%%%%%%%%%%
\subsection{Simulations}

Our simulations are those introduced in \cite{BAHAMAS_SIDM}, which combined the galaxy formation code BAHAMAS \citep{BAHAMAS} with the SIDM code used in \cite{SIDM_bullet}. They were run using a WMAP 9-yr cosmology\footnote{With $\Omega_\mathrm{m}=0.2793$, $\Omega_\mathrm{b}=0.0463$, $\Omega_\mathrm{\Lambda}=0.7207$, $\sigma_8 = 0.812$, $n_\mathrm{s} = 0.972$ and $h = 0.700$.} \citep{WMAP9}.

This paper uses simulations run with four different models of dark matter: CDM (i.e.\ zero self-interaction cross-section) plus SIDM0.1, SIDM0.3 and SIDM1 (which have velocity-independent cross-sections of $0.1$, $0.3$ and $1 \, \mathrm{cm^{2}/g}$ respectively). For each model, we have a $400 \, h^{-1} \mathrm{Mpc}$ box simulated with dark matter and baryon particle masses of $5.5 \times 10^{9} M_{\odot}$ and $1.1 \times 10^{9} M_{\odot}$ respectively. For CDM and SIDM1 we also have high-resolution simulations of a smaller volume, which we call CDM-hires and SIDM1-hires. The Plummer-equivalent gravitational softening length is 4$h^{-1}$kpc in physical coordinates below z = 3 and is fixed in comoving coordinates at higher redshifts. These have a box-size of $100 \, h^{-1} \mathrm{Mpc}$ and eight times better mass resolution than our standard resolution simulations.

The subgrid physics to model the baryonic prescription within the simulations was developed as part of the OWLS project \citep{OWLS}. Specifically {\sc BAHAMAS} includes radiative-cooling \citep{OWLScoolingRates}, star formation \citep{OWLSstarFormation}, stellar evolution and chemodynamics \citep{OWLSstellarEvoChemo} and stellar and AGN feedback \citep{OWLSagn,OWLSagn2}.

\subsection{Matching simulations to observations}
\label{subsect:samples}
In order to sample match those clusters in the suite of simulations and those used in H17 we must separate the relaxed clusters from dynamically unrelaxed. 
To do so we first take a random sample of 150 friends of friends (FOF) clusters with masses $10^{14}M_\odot<M_{\rm 200}<3\times10^{14}M_\odot$ and {\it all} clusters with $M_{\rm 200}>3\times10^{14}M_\odot$ over five different redshifts, $z=$0, 0.125, 0.250, 0.375 and 0.5. We choose this separation since there are very few large clusters, but many smaller ones which would computationally take too long to analyse. We then follow the same prescription as in H17 and take the ratio of the X-ray gas emission within 100kpc and 400kpc. This gives a proxy for how compact the X-ray gas is and in the case of relaxed halo with a cool core, this will be high. Studies show that this is good proxy for the dynamical state of a cluster with a cut at 0.2 as the divide between relaxed and disturbed \citep{dynamical_state_xray}. Table \ref{tab:data} gives the pre-cut and effective cluster members after we have made our selection. 

Having dynamically matched the two samples, we now extract the two components from the simulations; the dark matter and the stellar matter. To do this we run {\sc SExtractor} on the projected density distributions. We note here, that although this is not directly comparable to observations that use strong gravitational lensing, it does include many sources of error that are of importance. These include the projection effect of cluster members shifting the position, the physics associated with baryons and its coupling to dark matter, the inclusion of outliers that may be included in the sample, for example clusters that appear to be relaxed when in fact they have experienced recent mergers, and any bias due to cluster tri-axiality and small haloes close to the centre shifting the halo. Further aspects that are not captured by {\sc SExtractor} will be addressed in section \ref{sec:obs}.
  
%%%%%%%%%%%%%%%%%%%%%%%%%%%%%%%%%%%%%%%%%%%%%%%%%%%%%%%%%%%%
%%%%%%%%%%%%%%%%%%%%%%%%%%%%%%%%%%%%%%%%%%%%%%%%%%%%%%%%%%%%
\fig
\includegraphics[width=0.5\textwidth]{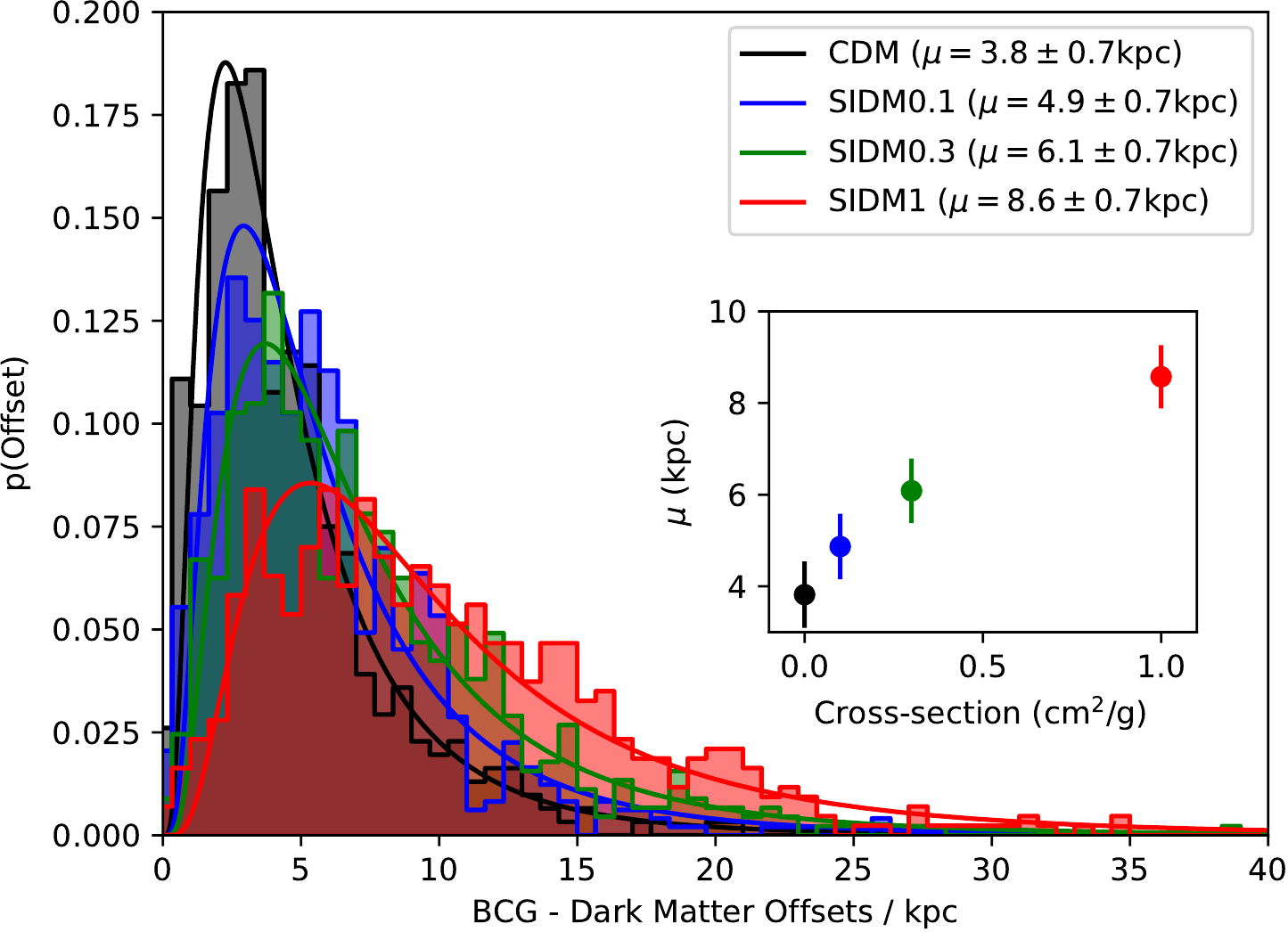}
\caption{\label{fig:resultsA} The complete sample of offsets between the brightest cluster galaxy and the dark matter halo for different cross-sections of dark matter. We fit log-normal distributions to each sample and report the median in the legend and the inset axis.}
\efig
%%%%%%%%%%%%%%%%%%%%%%%%%%%%%%%%%%%%%%%%%%%%%%%%%%%%%%%%%%%%

\section{Method}

K17 showed that the BCG of an SIDM galaxy cluster will oscillate in the gravitational potential of a cored density profile. The size of this oscillation should correlate with the core size and hence scale with cross-section. However, this signal is degenerate with the inherent measurement error associated with measuring the centre of a dark matter halo. H17 proposed a solution by modelling the distribution of DM-BCG offsets as the convolution of the distribution expected from a harmonic oscillator (with amplitude $A_{\rm w}$) with Gaussian measurement errors. To break the degeneracy between these two signals, H17 estimated the measurement errors from fitting NFW profiles to mock lensing data, generated using known NFW profile lenses. Using this they constrained $A_{\rm w}$.

Instead of attempting to break the degeneracy between measurement errors and genuine offsets, we note that the effect of SIDM is simply to broaden the distribution of DM-BCG offsets, whether it is wobbling or measurement error. We therefore choose to ignore the physical reason and merely measure the distributions from our simulations, add an additional noise component associated with strong lensing that H17 calculated empirically and then compare the final distributions with the observations.

In order to do this we first  infer the positions of the dark matter and the baryonic components using the peak finding algorithm SExtractor on the projected surface density map of each component. We then model the distribution of offsets between the dark matter halo and the BCG, $x$ with a log-normal probability density function, i.e.
\be
f(x) = \frac{1}{x \sigma  \sqrt{2\pi}} \, \exp\left( -\frac{(\ln x - \ln \mu)^2}{2\sigma^2} \right),
\ee
where $\mu$ is the median offset and $\sigma^2$ its logarithmic variance. We find the best fitting $\mu$ and $\sigma$ using the Maximum Likelihood Estimator function from Scipy\footnote{\url{https://docs.scipy.org/doc/scipy/reference/generated/scipy.stats.lognorm.html}}$^,$\footnote{\url{https://docs.scipy.org/doc/scipy/reference/generated/scipy.stats.rv_continuous.fit.html\#scipy.stats.rv_continuous.fit}}.

It has been noted in previous studies that using a particular algorithm to find the location of projected dark matter haloes can have an impact on the final result \citep{SIDM_bullet}. This was because in dynamically unrelaxed clusters can produce complex projected dark matter distributions, where different iso-density contours are centred on different points. As such, the position of the dark matter halo changes as a function of the scale on which the position is measured. Here we are dealing with relaxed clusters and therefore should not experience the same effect, however to test the sensitivity of our results to the choice of algorithm, we study how changing the size of the {\sc SExtractor}  kernel changes the distribution of offsets. For each simulated cross-section we take a sample of 150 clusters from the $z=0.25$ snapshot, with masses $M_{\rm 200}>3\times10^{14}M_\odot$. We then measure $\mu$ for a variety of different kernel sizes. We find that the best fitting $\mu$ is insensitive to the choice of kernel size and therefore we are confident that we are measuring an underlying trend and not an artefact of our estimator. Moreover the underlying trend should be independent of the choice of algorithm to find the halo centres. We choose to use a Gaussian kernel with a standard deviation of 9kpc for the rest of the paper

We now combine all of the offsets between the BCG and dark matter for each cross-section and measure their lognormal median and variance. Figure \ref{fig:resultsA} gives the resulting histograms with their best-fit log-normal distributions and the median of this in the legend and the inset axis.

We find that CDM has the smallest median of $\mu=3.8\pm0.7$ kpc, SIDM0.1, $\mu=4.9\pm0.7$ kpc, SIDM0.3, $\mu=6.1\pm0.7$ kpc and SIDM1, $\mu=8.6\pm0.7$ kpc. 

We find a strong correlation between $\mu$ and cross-section. However, if we are to infer the cross-section of dark matter from the observations, we must parametrise how the median offset depends on the cross-section. To do this we follow a forward modelling approach, whereby we take the simulations and apply know effects in order to produce a distribution that can be directly compared to simulations. We therefore state that the total, expected median offset that would be observed, $\mu_{TOT}$ is some function of cross-section, $\sigma/m$ plus some other unknown parameters, i.e.
\be
\mu_{\rm TOT} = g(\sigma/m, X).
\ee
where $X$ is a list of unknown parameters, which must be identified and then marginalised over. 
Here we identify three major concerns that will effect how we parameterise this function.
 \begin{enumerate}
  \item {\bf Finite resolution effects:} The results in Figure \ref{fig:resultsA}  are very close to the gravitational softening length of the simulations, where the gravitational forces become non-Newtonian. In bid to maximise the number of clusters available to the analysis, whilst minimising computational time, the chosen resolution was selected. However, on scales $r<10$kpc, effects could manifest themselves that impact the results. We therefore model any effects that the plummer softening length of the simulation, $\epsilon$ may have i.e.
 \be
\mu_{\rm TOT} = g(\sigma/m, \epsilon).
\ee
  \item {\bf Simulation analysis does not match that of the observations exactly}. The offset between the BCG and dark matter is a combination of the physical wobble and the inherent error in measuring the location of a dark matter halo with a constant density core. In order to compare the simulations directly to the observations, we must either forward model the simulations or deconvolve the expected error distribution from the observations. Given that we are attempting to forward model the simulations through $g$, we must incorporate the expected effect of observations on the median offset, $\mu$.
     \be
\mu_{\rm TOT} = g(\sigma/m, \epsilon,\hat{\sigma}),
\ee
where $\hat{\sigma}$ is an operator that will apply observational effects to the offset.
  \item {\bf Baryonic effects}. It has been recently shown that although more massive dark matter halos have larger cores (and hence expected to have larger median offsets), those that harbour a larger stellar mass will have a cuspier density profile \citep{CEAGLE_SIDM,SIDM_diversity}.
  As such, the concentration of stellar mass as well as the halo mass will likely impact the median offset given the scales in question and hence we must incorporate in to our final ansatz,
   \be
\mu_{\rm TOT} = g(\sigma/m, \epsilon,\hat{\sigma},M_{\rm 200}, M_\star).
\ee
The following sections will investigate each of these components further.

  %In H17 they quantified the observational error by simulating clusters with NFW density profiles. \cite{BAHAMAS_SIDM} found that the critical curves (and hence multiple image locations) change for different SIDM models. Therefore in order to quantify the true error in lensing position we would need to construct mock strong lensing maps and recover the dark matter and BCG positions. In this way we could completely fold in the observational error. We must 
%%%%%%%%%%%%%%%%%%%%%%%%%%%%%%%%%%%%%%%%%%%%%%%%%%%%%%%%%%%%
%%%%%%%%%%%%%%%%%%%%%%%%%%%%%%%%%%%%%%%%%%%%%%%%%%%%%%%%%%%%

% The BAHAMAS simulations have been tuned to have the correct total stellar to halo mass, however we find that the concentration of stellar mass on the scales of interest are somewhat underestimated. As such we model the dependency on the the stellar mass at these scales and taking estimates of the stellar mass from observations of CLASH cluster apply this to the observations. Therefore this requires accurate estimates of the stellar matter on these scales, which will be dependent on the chosen initial mass function plus many other parameters. Moreover, this model is based on the results of these simulations, and hence we rely on the relation between stellar mass and median offset to have converged on these scales. Investigating this is beyond the scope of this work, however, the model presented will need to be tested on larger scale, higher resolution simulations in the future.
  \end{enumerate}

%%%%%%%%%%%%%%%%%%%%%%%%%%%%%%%%%%%%%%%%%%%%%%%%%%%%%%%%%%%%
%%%%%%%%%%%%%%%%%%%%%%%%%%%%%%%%%%%%%%%%%%%%%%%%%%%%%%%%%%%
%%%%%%%%%%%%%%%%%%%%%%%%%%%%%%%%%%%%%%%%%%%%%%%%%%%%%%%%%%%%%
%\fig
%\includegraphics[width=0.5\textwidth]{Figures/WobbleFuncMass.pdf}
%\caption{\label{fig:resultsB} The median of the log-normal distribution $\mu$ as a function of mass and cross-section. Each bin has 45 clusters in them to ensure equal statistical power. We find for increasing cross-section both the intercept and gradient increase. CDM exhibits no such correlation with mass.}
%\efig
%%%%%%%%%%%%%%%%%%%%%%%%%%%%%%%%%%%%%%%%%%%%%%%%%%%%%%%%%%%%
%%%%
%%%%%%%%%%%%%%%%%%%%%%%%%%%%%%%%%%%%%%%%%%%%%%%%%%%%%%%%%%%%
%%%%%%%%%%%%%%%%%%%%%%%%%%%%%%%%%%%%%%%%%%%%%%%%%%%%%%%%%%%%
%\fig
%\includegraphics[width=0.5\textwidth]{/KernelSensitivity.pdf}
%\caption{\label{fig:KernelSize}  How sensitive SExtractor is to the choice of Kernel size. We measure the dark matter and stellar positions of each cluster using SExtractor with varying kernel sizes. We then estimate the median offset of the distribution. Here we show the results for a subset of haloes from each simulated cross-section. We find that the measured median is insensitive to the Kernel size and therefore measuring a true underlying trend and not an artefact of our choice of peak finding algorithm.}
%\efig
%%%%%%%%%%%%%%%%%%%%%%%%%%%%%%%%%%%%%%%%%%%%%%%%%%%%%%%%%%%%
%%%%%%%%%%%%%%%%%%%%%%%%%%%%%%%%%%%%%%%%%%%%%%%%%%%%%%%%%%%%
\subsection{Accounting for finite simulation resolution}\label{sec:resolution}
Our initial analysis of the simulations show that the expected median DM-BCG offset is $\mu\sim10$kpc and is therefore in proximity to the Plummer-equivalent gravitational softening  length of the simulations ($\epsilon=4h^{-1}$kpc) \citep{gadget2}. We therefore investigate how sensitive these results are to the resolution of the simulations. H17 found a significant difference between the low and high resolution simulations for CDM, and hence the $\sim4$kpc offset observed in CDM could be just the sensitivity limit of the simulation, which could also be impacting the other simulations. We therefore run two smaller, high-resolution boxes, one for CDM and one for SIDM1 and compare the predicted signals.

To do this, we first measure the best-fitting $\mu$ for the CDM and SIDM1-hires sample of $\sim20$ clusters, selected using the procedure described in Section~\ref{subsect:samples}. We then generate a mass-matched sample also of $\sim20$ clusters from the CDM and SIDM1 simulation, and measure $\mu$ for these samples. Given the large volume of the low resolution simulations, we can generate many such samples, and so we repeat this second step 300 times. Figure \ref{fig:testConvergence} shows the results. The red filled histograms show the measured distribution of $\mu$ from the 300 CDM (top panel) and SIDM1 (bottom panel) samples. The dotted vertical line and shaded region give the measured value and error from the high-resolution sample. We find that the high resolution simulations in both situations have lower medians compared to the low-resolution. 
 %%%%%%%%%%%%%%%%%%%%%%%%%%%%%%%%%%%%%%%%%%%%%%%%%%%%%%%%%%%%
%%%%%%%%%%%%%%%%%%%%%%%%%%%%%%%%%%%%%%%%%%%%%%%%%%%%%%%%%%%%
\fig
\includegraphics[width=0.5\textwidth]{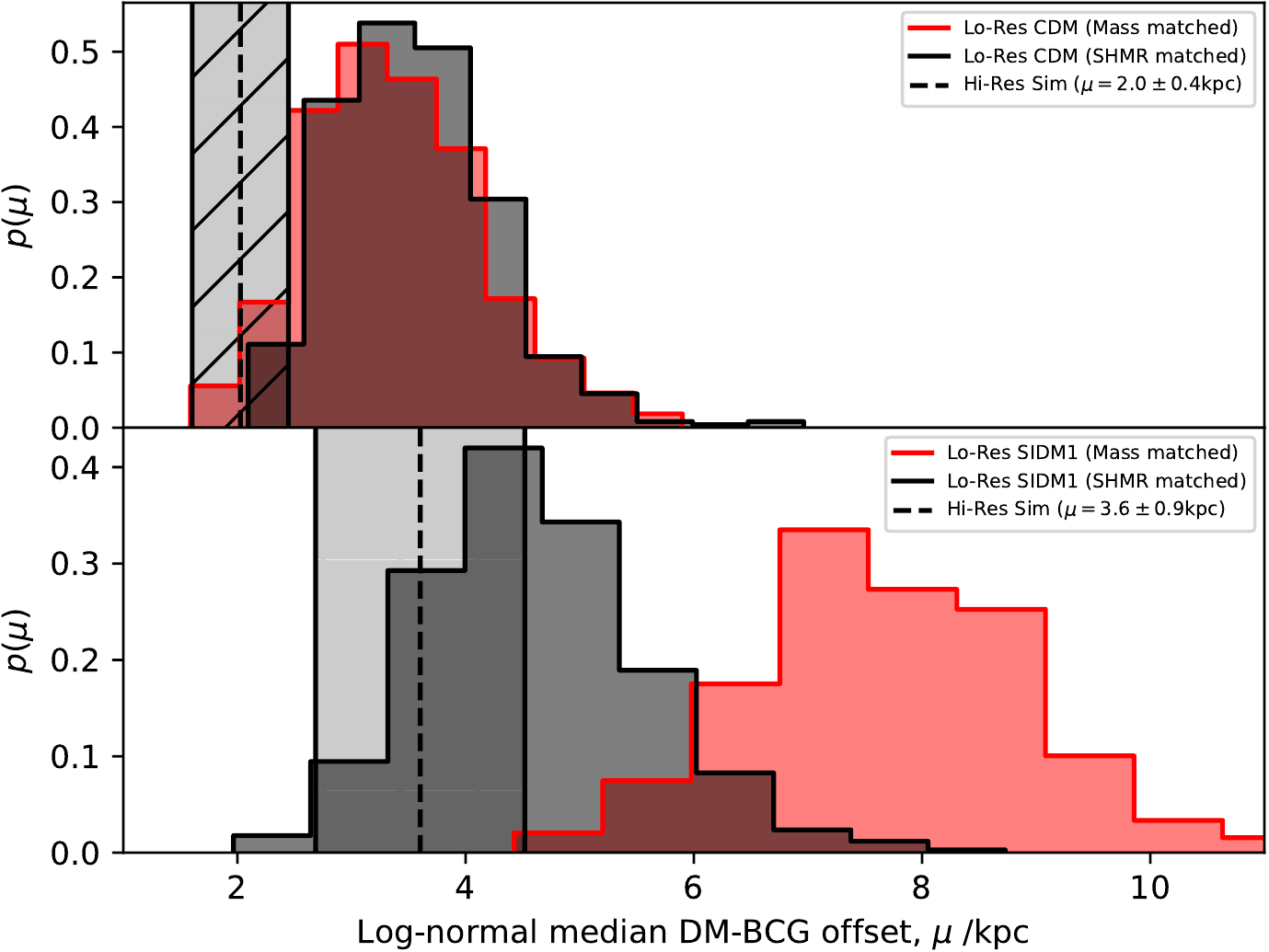}
\caption{\label{fig:testConvergence} Effect of finite resolution. We test whether high and low resolution simulations of the same simulation produce similar results given a similar mass distribution (red histogram) and sample size or similar stellar to halo mass ratio (black histogram). Given that the low resolution sample has many more haloes than the high-resolution we randomly sample the same number of clusters as in the high-resolution and measure $\mu$. The dotted vertical line shows the estimate from the hi-res simulation with the associated error-bar given by the shaded region. We find the estimated median in the high-resolution is under-predicted compared to that of the low-resolution and therefore must be modelled.}
\efig
%%%%%%%%%%%%%%%%%%%%%%%%%%%%%%%%%%%%%%%%%%%%%%%%%%%%%%%%%%%%

Looking closely at the density profiles of each sample, we find that the high-resolution haloes have denser stellar profiles than their low-resolution counterparts. In galaxy clusters with SIDM, denser stellar distributions lead to smaller dark matter cores \citep{CEAGLE_SIDM}, so in order to understand the differences due to {\it only} the resolution, (and not due to differences arising from different baryon distributions) we match the samples in stellar to halo mass ratio (SHMR) and re-calculate the distribution. The result is the black solid histograms in Figure \ref{fig:testConvergence}. We find that by matching the samples in SHMR, the agreement between the low and high-resolution is improved, however there remains some residual difference. We therefore apply very strict SHMR matching such that there are equal number of clusters in each low and high resolution sample and model the effect of the softening via the ansatz
\be
\mu_{\rm MEAS} = (\mu_{\rm SIM}^\gamma+(\alpha\epsilon)^\gamma)^{\frac{1}{\gamma}},
\label{eqn:conv}
\ee
where $\mu_{\rm MEAS}$ and $\mu_{\rm SIM}$ are the measured and intrinsic log-normal medians for a particular cross-section, and $\epsilon$ is the softening length of the simulation. 
Using two different resolution simulations, from two different cross-sections, (i.e. high and low res for SIDM1 and CDM), we are able to fit for the four parameters, $\gamma$, $\alpha$, $\mu_{\rm SIM, CDM}$ and $\mu_{\rm SIM, SIDM1}$. Once we have found $\gamma$ and $\alpha$, we are able to calculate $\mu_{\rm SIM}$ for any low or high resolution simulation (assuming that these values are constant for other cross-sections and halo masses).
%We use equation \eqref{eqn:conv} to approximately remove the effect of finite resolution and present the distributions of $\mu_{\rm T}$ in the left panel of Figure \ref{fig:WobbleFuncMass}. 

%%%%%%%%%%%%%%%%%%%%%%%%%%%%%%%%%%%%%%%%%%%%%%%%%%%%%%%%%%%%
%\begin{table}
%\centering
%\begin{tabular}{|c|c|c|}
%Sample & $X_1 $/ kpc  & $X_2$ / kpc \\
%\hline
%CDM & $4.04 \pm 0.25$ & $-0.49 \pm 0.52$ \\
%SIDM0.1 & $4.81 \pm 0.23$ & $0.50 \pm 0.66$ \\
%SIDM0.3 & $4.92 \pm 0.34$ & $3.26 \pm 0.73$ \\
%SIDM1 & $7.31 \pm 0.29$ & $3.55 \pm 0.61$ \\
%\hline
%\end{tabular}
%\caption{ Fitted coefficients assuming a linear correlation between the median of the log-normal distribution $\mu$ and the log of the halo mass.
%\label{tab:MassFuncCoeffs}}
%\end{table}
%%%%%%%%%%%%%%%%%%%%%%%%%%%%%%%%%%%%%%%%%%%%%%%%%%%%%%%%%%%%

%%%%%%%%%%%%%%%%%%%%%%%%%%%%%%%%%%%%%%%%%%%%%%%%%%%%%%%%%%%%
   \figs
\includegraphics[width=\textwidth]{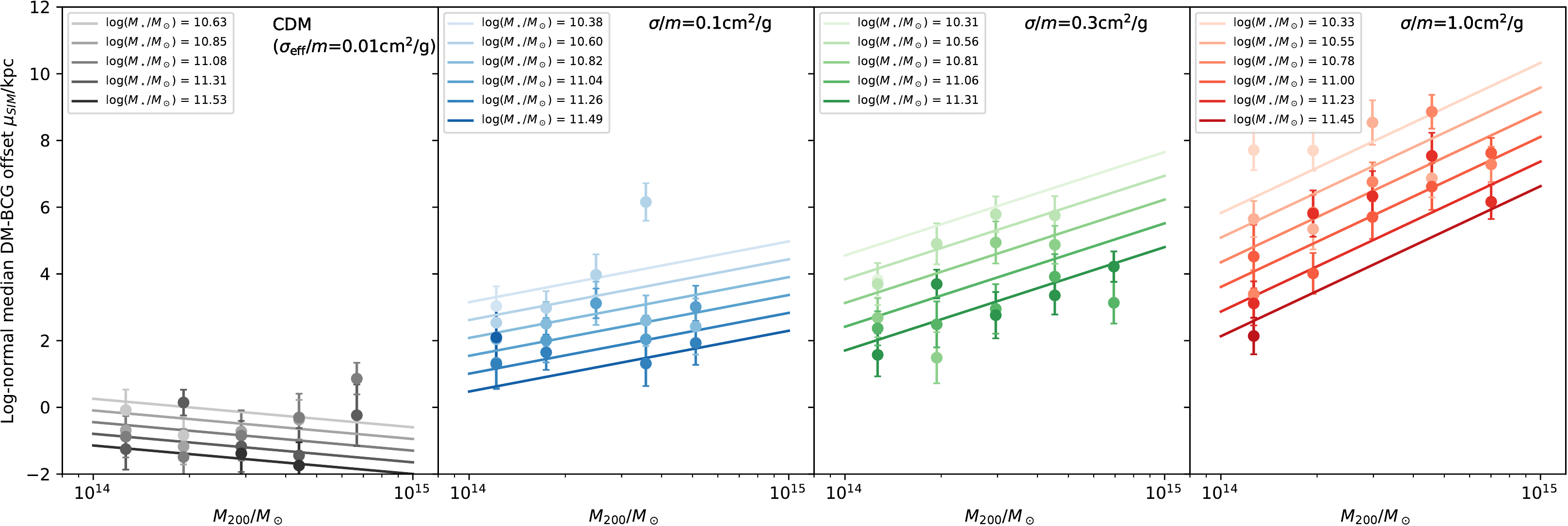}
\includegraphics[width=\textwidth]{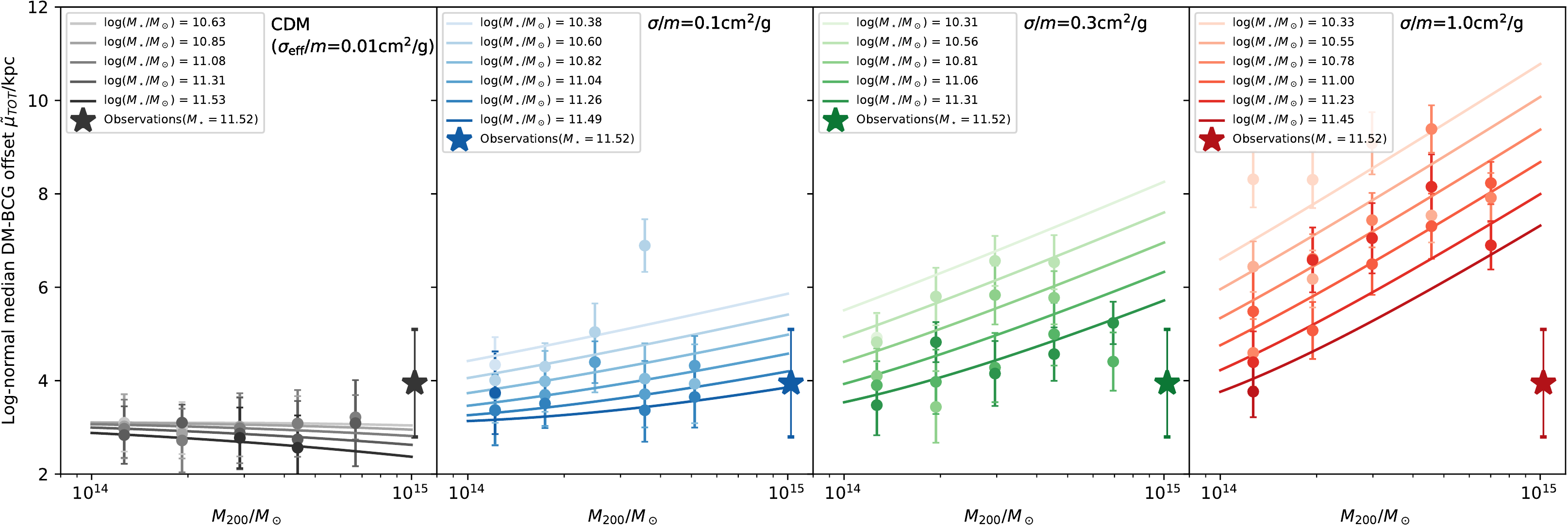}
\caption{\label{fig:SIDMmodel} 
Our final model for the observed offset between the BCG and dark matter in relaxed clusters, as a function of the halo mass, stellar mass within $10$kpc (different lines) and self-interaction cross-section of dark matter, $\sigma_{\rm DM}/m=0, 0.1, 0.3, 1.0$ from left to right, in the absence of any experimental noise ({\it Top}) and with observational noise ({\it Bottom}). For comparison we show the observations in the bottom row with a star and the color representing the stellar mass (also shown in the legend). Since we assume a model in log$\sigma$, where CDM is ill-defined, we show the effective cross-section of our model. This represents the validity limit $\sigma/m=0.01$cm$^2$/g.}
\efigs
%%%%%%%%%%%%%%%%%%%%%%%%%%%%%%%%%%%%%%%%%%%%%%%%%%%%%%%%%%%%

  \subsection{Applying observational effects}\label{sec:obs}
In order to fully forward model the simulations in order to directly compare with observations we must add an additional source of error.
However carrying out a full mock gravitational lensing analysis on the simulated clusters is beyond the scope of this paper and therefore we choose to numerically modify $\mu_{\rm SIM}$. 

Instead we convolve the effect of observational noise on to the simulation data by numerically adding random Gaussian noise to each radial offset in the measured log-normal distributions. We then re-measure the log-normal distributions. Since we do this numerically, to get accurate results it takes some time. Therefore in order to speed this up we test whether this numerical method has a analytical form. Given that in all sense this is just a convolution of a log-normal radial and delta function with a Gaussian distribution, it is not possible to analytical calculate. We therefore carry out some mocks tests with a known log-normal, add on the observational noise and re-calculate the median. As figure \ref{fig:convolutionEffect} shows, we find that the resulting median is almost exactly the sum of the original median and the width of the Gaussian, added in quadrature, i.e.
\fig
\includegraphics[width=0.49\textwidth]{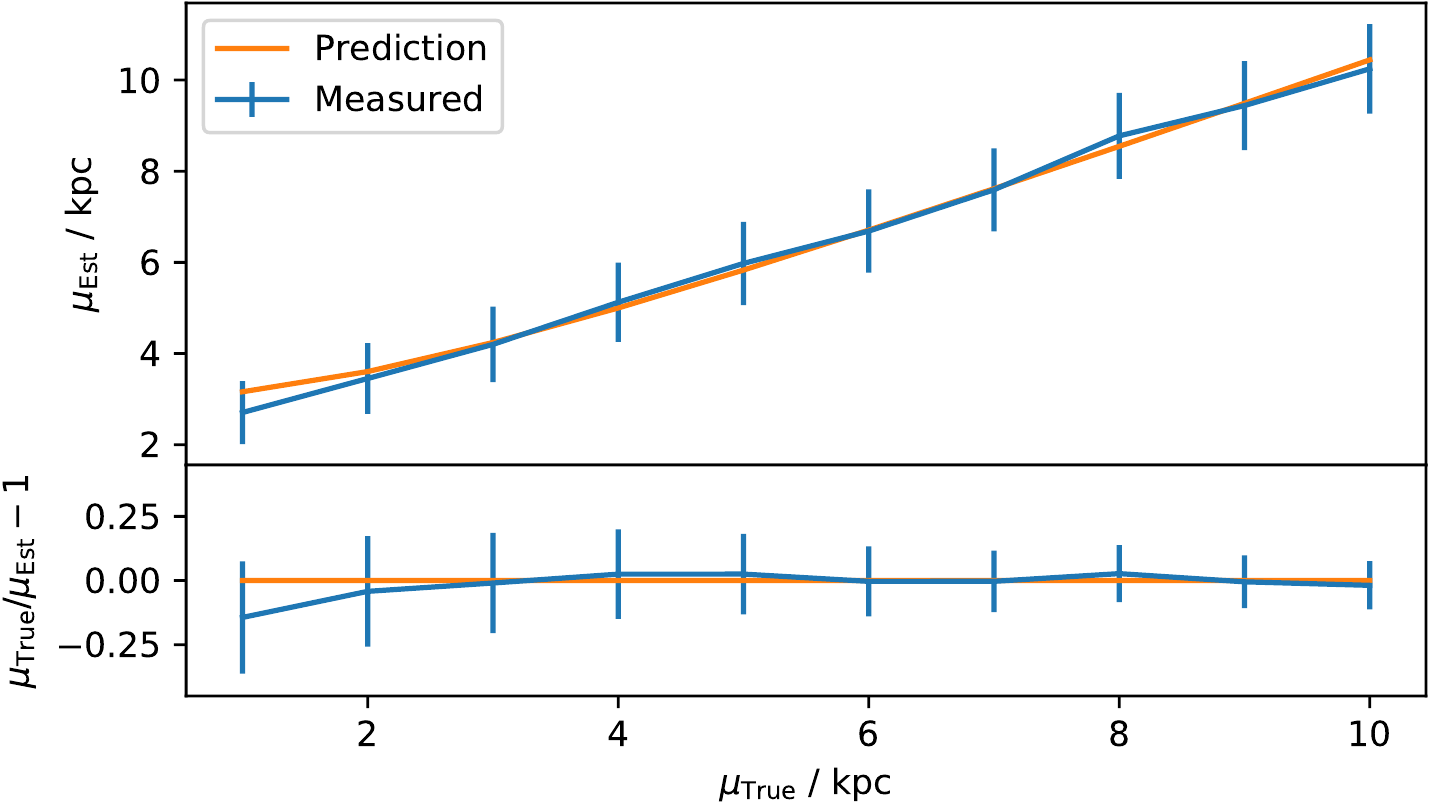}
\caption{\label{fig:convolutionEffect} The effect of convolving a log-normal distribution with a Gaussian of width $\sigma_{\rm obs}=3.1$kpc. The prediction is the simply the $\mu_{\rm SIM}$ added in quadrature with the observational noise. }
\efig
We therefore choose to model the effect of observational noise on the median offsets by adding them in quadrature, i.e.
\be
\mu_{\rm TOT}^2 = \mu_{SIM}^2 + \sigma_{\rm obs}^2.
\ee
   
%%%%%%%%%%%%%%%%%%%%%%%%%%%%%%%%%%%%%%%%%%%%%%%%%%%%%%%%%%%%
\begin{table}
\centering
\begin{tabular}{|c|c|c|c|c|c|}
Sample & $\mu_{\rm MEAS, lo}$ (kpc ) &$\mu_{\rm MEAS,hi}$ (kpc )&$\mu_{\rm SIM}$ (kpc ) \\
\hline CDM &  $3.80 \pm 0.7$ & $2.0 \pm 0.4$ & $0.8^{+0.9}_{-0.8}$  \\
 SIDM1 &  $5.0 \pm 1.0$ & $3.6 \pm 0.9$ &$2.3^{+1.8}_{-0.7}$ \\
\hline
\end{tabular}
\caption{ The fitted parameters to equation \eqref{eqn:conv}, where we model the effect of the finite resolution of the simulation on our results. We also find that $\alpha=0.41^{+0.37}_{-0.01}$ and $\log_{10}(\gamma)=-0.02^{+0.61}_{-0.01}$.
\label{tab:conv}}
\end{table}
%%%%%%%%%%%%%%%%%%%%%%%%%%%%%%%%%%%%%%%%%%%%%%%%%%%%%%%%%%%%
\subsection{The impact of baryons on the dark matter}
Initial studies clearly showed the expected offsets were well within the stellar distribution of the BCG. Thus sub-grid physics models that may affect the distribution of stellar matter will likely impact the signal we observe. This hypothesis was backed up when we noticed the difference in expect median offset between a mass matched sample and a sample matched in SHMR in the previous section.

How the baryons impact our results will primarily depend on how well the BAHAMAS simulation do at reproducing the observed stellar mass distribution.
The BAHAMAS simulations have been tuned to return the correct stellar mass function, and as such the observed H17 cluster sample SHMR relation {\it should} match the simulated one. If this is the case then constraints derived from a representative sample should be unbiased. Using stellar masses from \cite{CLASHbcgMass} we find that indeed this is the case, and the SHMR of the observed clusters well matches the simulated ones. 

However, the stellar mass is measured up to $\sim50$kpc, well beyond the scales that are probed by this technique, which are closer to $\sim10$kpc. This is particularly important since although the simulations have been tuned to give the correct total stellar mass, the amount of stellar mass on the scales in question could be very different. As such we look closer at the distribution of stellar mass within $\sim10$kpc. Using estimates from \cite{CLASHbcgDensity} we find that in fact the observed clusters have a much high density of stellar mass within $10$kpc. As such we are motivated to model the behaviour of the median offset as a function of halo mass {\it and} stellar mass within $10$kpc. 

  %However following our investigations in to the dark matter - stellar matter coupling, it is insufficient to simply use the halo mass. We therefore must estimate the stellar mass within $10$kpc of the observed clusters. To facilitate this we adopt the stellar mass concentrations from \cite{CLASHbcgDensity}, which look at predominantly CLASH clusters (where the majority of H17's sample is sourced from). We apply the transformation as before in order to compare to observations and show in the bottom right hand panel of Figure ]\ref{fig:WobbleFuncMass} where the stellar mass within $10$kpc for the observed sample lies. We see that the sample is much denser than the  majority of the halos in the simulations, and as such it relaxes our conclusions concerning constraints on the cross-section.

\be
\mu_{\rm SIM} = X_1 +  X_2\log_{10}\left(\frac{ M_{200}}{10^{14}M_\odot}\right) + X_3\log_{10}\left(\frac{ M_\star(<10{\rm kpc})}{10^{11}M_\odot}\right),
\label{eqn:muMassSIDM}
\ee
 where the relation to cross-section could be either 
\be
X_{i}(\sigma) = a_i+b_i\log_{10}\left(\frac{\sigma/m}{1{\rm cm}^2/{\rm g}}\right) {\rm~for~} i=1,2,
\label{eqn:logSIDM}
\ee
or
\be
X_{i}(\sigma) = a_i+b_i\left(\frac{\sigma/m}{1{\rm cm}^2/{\rm g}}\right) {\rm~for~} i=1,2.
\label{eqn:linSIDM}
\ee
%Given that CDM ($\sigma^2 =0$) is not defined in this logarithmic model, we only fit equation \eqref{eqn:logSIDM} to the three finite cross-sections. However, we report the cross-section that the CDM simulations lie at, which effectively is the sensitivity limit of these simulations. 

\fig
\includegraphics[width=0.49\textwidth]{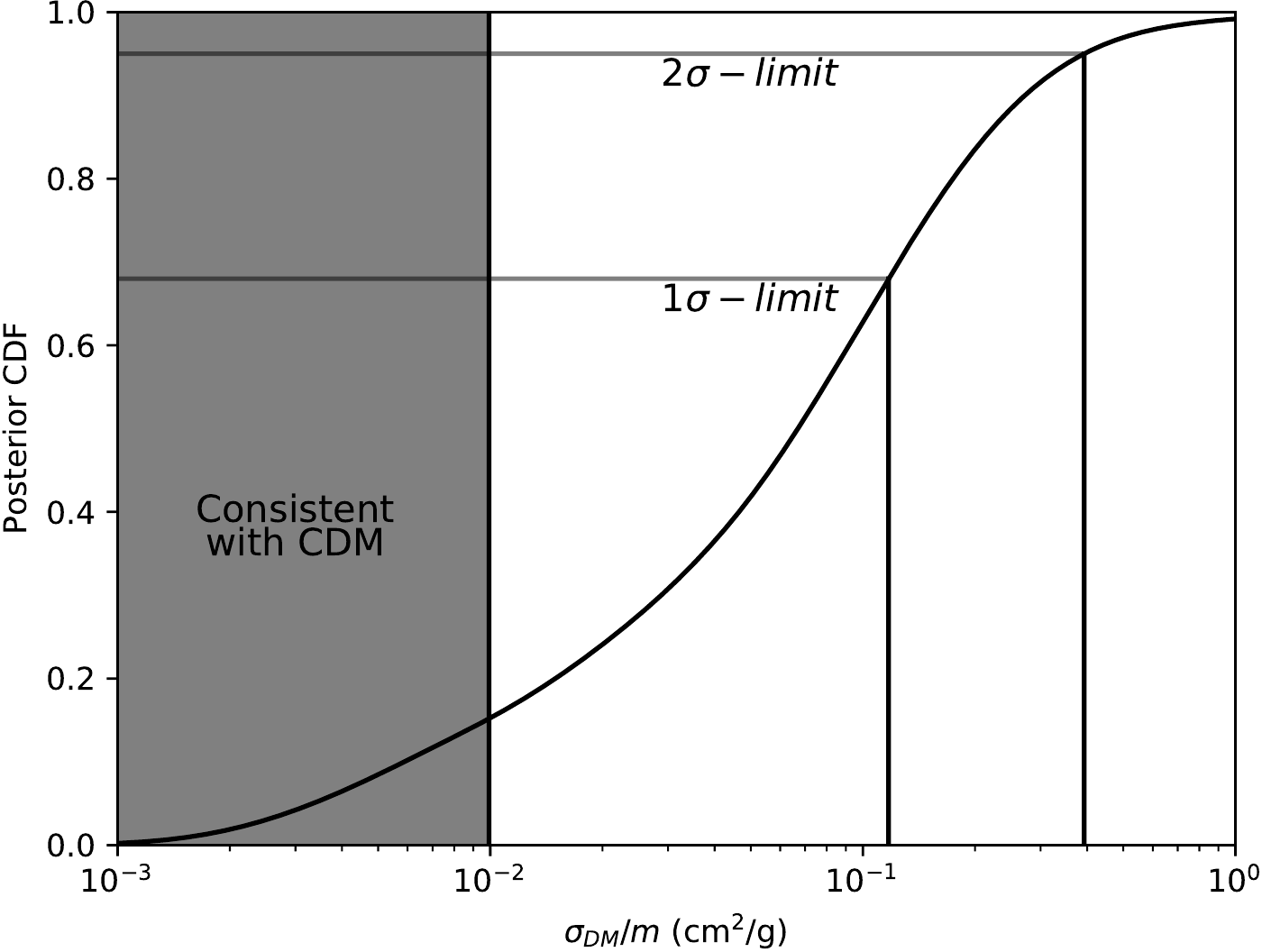}
\caption{\label{fig:results} 
The cumulative probability distribution of the observations given our model (assuming equation \eqref{eqn:logSIDM}), folding in errors associated with the parameters in equation \eqref{eqn:conv}. We give the one and two sigma limits of the observations which corresponds to $\sigma_{\rm DM}/m < 0.12 (0.39)$ cm$^2$/g at 68\% (95\%) confidence limit. Given that in this model CDM ($\sigma/m=0$cm$^2$/g) is undefined, we estimate the validity limit of the model, and the sensitivity limit of the simulations. The CDM offsets can be interpreted as $\sigma_{\rm CDM}=0.01$cm$^2$/g and is given by the shaded region. We find that 15\% of the probability of the observations lie in this region, hence showing that the observations are in tension with CDM at the $\sim1.5\sigma$ level.}
\efig

Following this we carry out the second fit using a least squares and a modified loss function to determine $a_i$ and $b_i$;
\be
\chi^2 = \sum_{i=0}^{nSim} \frac{ (\tilde{\mu}_{SIM_i} - \mu_{SIM_i})^2}{\sigma_{SIM_i}},
\label{eqn:lossFun}
\ee
where $\tilde{\mu}_{SIM_i}$ is the model median value and ${\mu}_{SIM_i}$ is the actual measured median value for the $i$th cross-section. In the case where we assume a log-cross-section ansatz, we sum over only finite cross-sections, whereas for a model linear in cross-section we include also CDM.

In order to choose between a linear or log cross-section model,  we compute the Bayesian information criterion (BIC), which penalises any good fit by the number of parameters used in the model in an attempt to reduce overfitting. The BIC can be computed by
\be
{\rm BIC} = -2\ln(L)+k\ln(n),
\ee
where L is the likelihood of the maximised model, $n$ is the total number of data points and $k$ is the number of parameters.
We give the corresponding BIC for each model in the fifth column of Table \ref{tab:loglinPars}. We find $\Delta$BIC$=4$ between the two models, which corresponds to a preference for the log-model. We therefore adopt this model and show it in Figure \ref{fig:SIDMmodel} with individual models in the right hand panel. %We also show the observational estimate from H17 in magenta. To aid the reader, the solid black lines give the model prediction for the three simulated finite cross-sections, ( $\sigma_{\rm DM}/m = 0.1$, $0.3$ \& $1.0 $cm$^2$/g.). We also show the fits from the right hand-panel of Figure \ref{fig:WobbleFuncMass}. 
\begin{table*}
\centering
\begin{tabular}{|c|c|c|c|c|c|c|c|}
Model & $a_{1}$ &$b_{1}$ &$a_2$  & $b_2$ & $a_3$ & $b_3$ &BIC\\
 & [ kpc ] & [ kpc ] & [ kpc ] & [ kpc ] & [ kpc ] & [ kpc ] &  \\
\hline 
Log&  $ 4.5 \pm 0.4 $&  $ 2.7 \pm 0.8 $&  $ 3.6 \pm 0.2 $&  $ 2.0 \pm 0.4 $&  $ -3.3 \pm 0.4 $&  $ -0.9 \pm 0.7 $ & 62 \\
Linear&  $ 1.9 \pm 0.4 $&  $ 2.6 \pm 0.7 $&  $ 1.4 \pm 0.2 $&  $ 2.4 \pm 0.3 $&  $ -2.1 \pm 0.3 $&  $ -1.3 \pm 0.6 $ & 66 \\
\hline
\end{tabular}
\caption{ The fitted coefficients to equation \eqref{eqn:logSIDM} and \eqref{eqn:linSIDM}, where we model the BCG-dark matter offset as a function of mass, stellar mass and cross-section. The sixth column shows the Bayesian information criterion that allows a comparison of the two models. With a  $\Delta$BIC$=4$ there is a preference for a log cross-section model.
\label{tab:loglinPars}}
\end{table*}
%%%%%%%%%%%%%%%%%%%%%%%%%%%%%%%%%%%%%%%%%%%%%%%%%%%%%%%%%%%%
 
\section{Results}{\label{sec:results}
Following the construction of our model, we have a total of 10 {\it non}-independent parameters;
\be
\theta = \{ \mu_{SIM, CDM},\mu_{SIM, SIDM1}, \alpha, \gamma, a_1, b_1, a_2, b_2, a_3, b_3 \}.
\ee
They are not independent since the estimate of the softening model parameters will effect the following $a_i$ and $b_i$.
As such we carry out the fit to the four softening model parameters first using a least-squares algorithm. Table \ref{tab:conv} gives the results of the fit.

We show the final model in Figure \ref{fig:results}. The top row shows the best fit model for the three finite cross-sections. In each panel we show the model estimate $\tilde{\mu}_{\rm SIM}$ as the solid lines and the actual measured estimate $\mu_{\rm SIM}$ as the data points to which the model is fitted, as a function of halo mass and stellar mass within $10$kpc. The bottom row shows the total, expected model, $\mu_{\rm TOT}$ after adding observational noise. We estimate the median offset of the H17 sample of clusters  finding  that $\mu_{\rm obs} = 3.9\pm1.2$kpc. We show this estimate as the star in each panel, where the colour of the star and legend gives the estimated stellar mass. 
The corresponding model parameters can be found in Table \ref{tab:loglinPars}.

Given that CDM ($\sigma_{\rm DM}/m =0$) is not defined in this logarithmic model, however gives a finite offset, we calculate what the effective cross-section the offsets predict in this model. This cross-section represents the sensitivity and validity limit of the simulations. We find that the effective cross-section of CDM is $\sigma_{\rm DM}/m=0.01\pm0.007$ cm$^2$/g.

\subsection{Constraints on the self-interaction cross-section}
 We now use the fitted models to directly constrain the cross-section of dark matter. In order to do this we must fold in the uncertainties of our model mainly driven by the softening model, since all subsequent parameters are derived from these. To do this we carry out the following prescription:
 \begin{enumerate}
 \item We first draw a sample randomly from the estimates of $\mu_{\rm SIM,lo}$ and $\mu_{\rm SIM,hi}$ (from Table \ref{tab:conv}), sampling from Gaussian distributions centred on the quoted means and with widths (standard deviations) given by the quoted errors. 
 \item From these estimates, we re-derive the four softening-model parameters (equation \eqref{eqn:conv}).
 \item Using the newly generated softening model we re-fit for $\mu_{\rm SIM}$ via equations \eqref{eqn:muMassSIDM} and equation \eqref{eqn:lossFun} and add observational noise to get a model of $\mu_{\rm TOT}$.
 \item Assuming a Gaussian probability density distribution in $\mu_{\rm obs}$, we calculate the cumulative density distribution (CDFs) in $\sigma/m$ using our new model of $\mu_{\rm TOT}$, adopting stellar mass estimates from \cite{CLASHbcgDensity}.
 \item Repeating $10^3$ times, we generate multiple CDFs and then take the mean to get the final CDF.
 \end{enumerate}
 The final mean CDF can be found in Figure \ref{fig:results} corresponding to a upper limit of   $\sigma_{\rm DM}/m < 0.12~(0.39)~$cm$^2$/g 68\% (95\%). We find that  15\% of the probability lies below the sensitivity threshold of the simulations ( $\sigma_{\rm DM}/m < 0.01$cm$^2$/g ), and is therefore consistent with CDM. This limit is illustrated by the shaded region.

 \subsection{Future prospects}
 This study has shown that with only a small number of strong lensing galaxy clusters we are able to place tight constraints on the self-interaction cross-section of dark matter. With future studies soon to come online we investigate how this method scales statistically. To this end we calculate the predicted 95\% constraints for two future studies: SuperBIT, a balloon-borne telescope that will image 200 galaxy clusters \citep{superbit}, and Euclid \citep{EUCLID}, a space-based telescope that will image $\sim10^{3}-10^{5}$ clusters. We calculate the constraints as a function of the average error in a single cluster, $\sigma_{\rm obs}$. To do this we take the H17 value and error and reduce the error by a factor of $\sqrt{N_{\rm cl}}$, and shift the median, $\mu$ for different values of $\sigma_{\rm obs}$. Figure \ref{fig:forecasts} shows the results. Each dotted line is a study with increasing precision on a single cluster. The solid cyan line is the sensitivity of this study, and the sensitivity regions of each survey are given in pink (cyan) for SuperBIT (Euclid). We find that, although an initial increase in sample size will improve the constraints by a factor of $\sim2$, further improvements would be moderate. As such, this experiment would be ideal for a survey the size of SuperBIT. A precision on dark matter astrometry of $\sim10$kpc, which can be achieved with weak gravitational lensing, will place discriminatory constraints and therefore could be of interest in the future. 
 
  \fig
\includegraphics[width=0.49\textwidth]{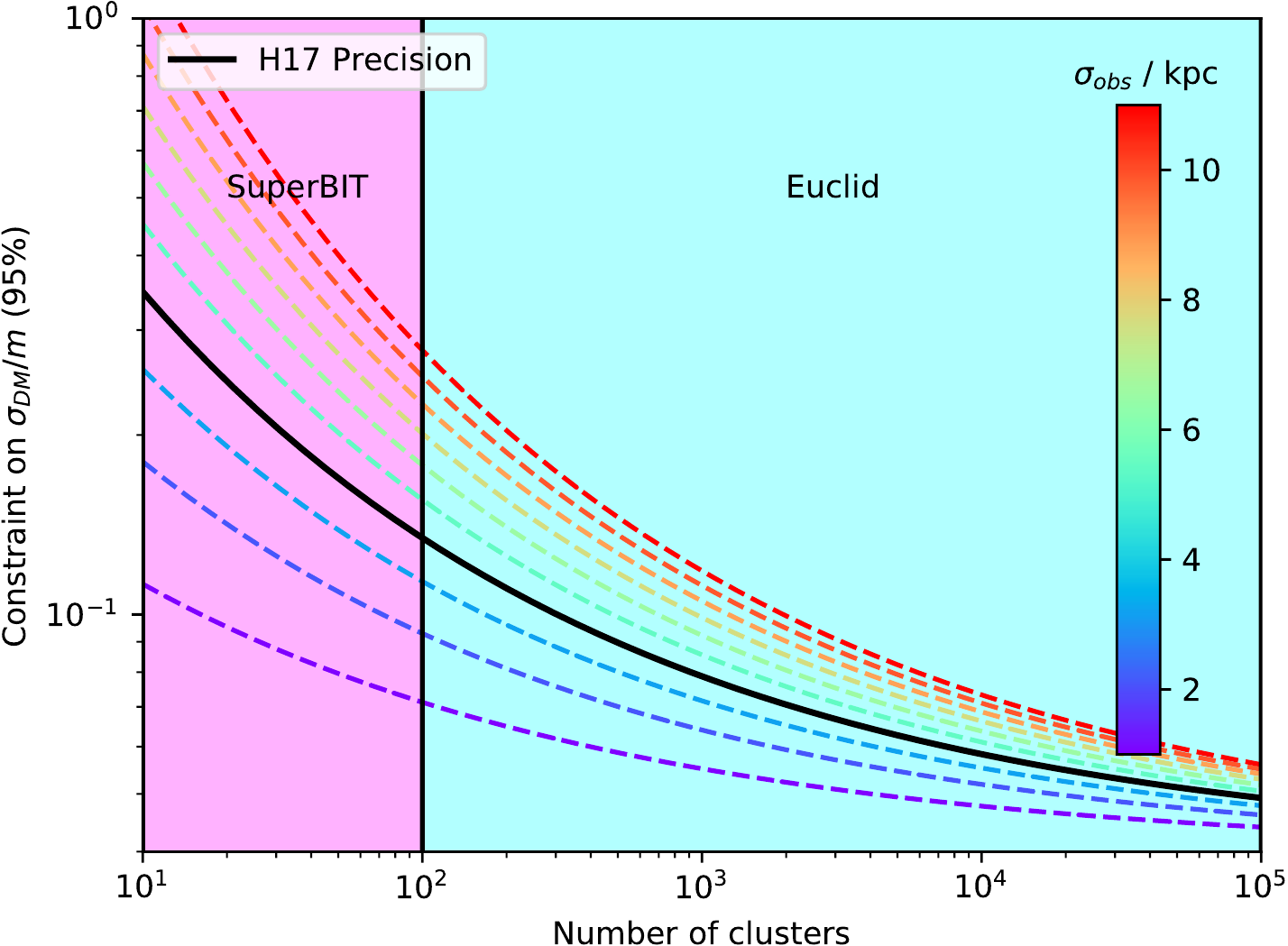}
\caption{\label{fig:forecasts} 
 Forecasted 95\% confidence limits for future surveys as a function of the number of clusters and the precision of a single cluster estimate. SuperBIT and Euclid will observe of order $10^2$ and 10$^{5}$ clusters respectively. The solid black line is the precision found in H17, for observations of strong gravitational lensing by 10 clusters. We find that although SuperBIT will yield a factor $\sim2$ improvement, large surveys like Euclid will only bring diminishing returns. Interestingly, even weak lensing observations with a precision on dark matter astrometry of only $\sim10$kpc may be able to place discriminating constraints in the future.}
\efig
\section{Discussion \& Conclusions}{\label{sec:conc}
We have used cosmological simulations of cold dark matter and self-interacting dark matter that include realistic baryonic physics to constrain the velocity independent, elastic, self-interaction cross-section of dark matter. 

It is predicted that during the collision of two galaxy clusters that harbour cored density profiles, the Brightest Cluster Galaxy (BCG) will be initially offset from the centre of the halo. Long after the relaxation of the cluster, this offset can persist with the BCG tracing out the motion of a harmonic oscillator \citep{darkgiants}. In CDM, the central density profile is cuspy and hence the BCG will be bound tight to the centre of the dark matter halo, however in models of dark matter that predict cores this will be a clear signal for a non-standard model of dark matter.  

In a recent paper, the distribution of offsets between the BCG and dark matter halo was estimated in 10 galaxy clusters. Fitting a two component model they estimated that the wobble amplitude, $A_{\rm w}\sim11$kpc \citep{Harvey_BCG}, in close agreement with previous studies \citep{densityProf2}. They compared this to high resolution simulations of CDM and found a discrepancy at the $\sim3\sigma$ level.

In this paper we have extended this comparison to include simulations with velocity-independent dark matter self-interactions. 
The simulations were run with four different cross-sections: $\sigma_{\rm DM}/m = 0$ (CDM), $0.1$, $0.3$ and $1.0$ cm$^2$/g. Modelling the distribution of BCG-DM offsets as a log-normal, we found that the median offset, $\mu$, increased with cross-section: 
$\mu_{\rm CDM}=3.8\pm0.7$ kpc, 
$\mu_{\rm 0.1}=4.9\pm0.7$ kpc, 
$\mu_{\rm 0.3}=6.1\pm0.7$ kpc and 
$\mu_{\rm 1.0}=8.6\pm0.7$ kpc. 

In order to infer the cross-section of dark matter from the simulations we construct a model that relates the median offset to the cross-section. We identify three clear concerns that are folded into this model;
\begin{enumerate}
\item The effect of the close proximity of the signal to the finite smoothing length of the simulation
\item The effect of observational noise on the signal
\item The impact of baryons in the core of the cluster
\end{enumerate}
Parameterising each, we estimate the final cross-section of dark matter, finding that   $\sigma_{\rm DM}/m < 0.12~(0.39)~$cm$^2$/g 68\% (95\%). Under the assumption that the model scales with the log of the cross-section, CDM is undefined. We therefore use the CDM simulations to estimate the validity limit of this model and the sensitivity limit of the simulations. We find that the offsets observed in CDM are interpreted as an effective cross-section of $\sigma/m=0.01$cm$^2$/g. Given our observations, we find that 15\% of the probability lies within this region and hence the observations are consistent with CDM to within $\sim1.5\sigma$.

The consequence of this limit is that models of SIDM that can significantly alter the structure of dwarf galaxy dark matter haloes, would require a cross-section that varies with the relative velocity between dark matter particles.
With observations of just 10 galaxy clusters, this method is almost at the precision necessary to discriminate between (and potentially rule out) otherwise viable models of dark matter. Future surveys, such as observations by SuperBIT of weak lensing around $\sim200$ clusters, will soon have the power to make dramatic impact.

\section*{Acknowledgments}
This research is supported by the Swiss National Science Foundation (SNSF). DH also acknowledges support by the Merac foundation and the ITP Delta foundation.
AR is supported by the European Research Council (ERC-StG-716532-PUNCA) and the UK STFC (ST/N001494/1).
RM is supported by a Royal Society University Research Fellowship.
This project has received funding from the European Research Council (ERC) under the European Union's Horizon 2020 research and innovation programme (grant agreement No 769130).
We would also like to thank Joop Schaye for his contributions to the BAHAMAS simulations.
This work used the DiRAC Data Centric system at Durham University,
operated by the Institute for Computational Cosmology on behalf of the
STFC DiRAC HPC Facility (www.dirac.ac.uk). This equipment was funded by
BIS National E-infrastructure capital grant ST/K00042X/1, STFC capital
grants ST/H008519/1 and ST/K00087X/1, STFC DiRAC Operations grant
ST/K003267/1 and Durham University. DiRAC is part of the National
E-Infrastructure.

\bibliographystyle{mn2e}
\bibliography{bibliography}

\bsp

\label{lastpage}

\end{document}